\documentstyle[12pt]{article}
\begin{document}
\parskip 10pt plus 1pt
\title{CONSTRUCTION OF MAXIMALLY SYMMETRIC SOLUTIONS FOR THE METRIC}
\date{}
\maketitle
\centerline{\it Debashis  Gangopadhyay\footnote{S.N.Bose National Centre For 
Basic Sciences,JD-Block,Sector-III,Salt Lake,Calcutta-700091,INDIA.
e-mail:debashis@bose.ernet.in}  
and   Soumitra Sengupta\footnote{Department of Physics,Jadavpur University,
Calcutta-700032,INDIA} }
\baselineskip=20pt

\newpage
\begin{abstract}
We construct new maximally symmetric solutions for the metric.We then show 
that for a string moving in a background consisting of maximally symmetric
gravity, dilaton field and second rank antisymmetric tensor field,the
O(d) $\otimes$ O(d) transformation on the vacuum solutions, in general, gives
inequivalent solutions that are {\it not} maximally symmetric.
\end{abstract}

\newpage

Some time back it was shown that the low energy string effective
action possesses, for time dependent metric $G_{\mu\nu}$, torsion
$B_{\mu\nu}$  and dilaton $\Phi$ background fields $(\mu, \nu =
1, 2, ..d)$ a full continuous O(d,d) symmetry under which
"cosmological" solutions of the equations of motion are
transformed into other inequivalent solutions [1]. Subsequently,
a generalisation to this was obtained [2]. These transformations
are conjectured to be a generalisation of the Narain construction
[3] to curved backgrounds.

Here we investigate the consequences of this O(d) $\otimes$ O(d)
transformation on the space-time symmetries of the theory. We
consider a string propagating in a gravity, dilaton and second
rank antisymmetric tensor background and show that {\it if the full
metric corresponding to a given background is maximally symmetric
then under the O(d) $\otimes$ O(d) twist this symmetry is not
preserved}.

We first discuss the meaning of maximal symmetry and the 
O(d) $\otimes$ O(d) symmetry.We then show that for $B_{\mu\nu}=0$, the only
maximally symmetric solution is that for which the dilaton background is a 
constant and the curvature constant $K=0$.This maximal symmetry is broken 
under the O(d)$\otimes$ O(d) transformation. We then show that an approximate 
maximally symmetric solution with $B_{\mu\nu} \not=0$ and non-zero curvature
is possible with a linear dilaton background.However, this symmetry is again
destroyed under O(d)$\otimes$ O(d) twist. 

For a maximally symmetric space-time [4] the curvature
$$R_{iklm} = K\left(g_{im}\enskip g_{kl} - g_{il}\enskip g_{km}\right)
\eqno(1)$$
K is the curvature constant proportional to the scalar curvature
$R^{i}_{i}$. Two maximally symmetric metrics with the same K and
the same number of eigenvalues of each sign, are related by a
coordinate diffeomorphism [4]. Now consider the low energy
effective action of string theory in D space-time dimensions.
This is [5] :
$$S = - \int d^{D}X \sqrt{det G} e^{-\phi} \left[\Lambda -
R^{(D)} (G) + \left({1 \over 12}\right) H_{\mu\nu\rho}H^{\mu\nu\rho} -
G^{\mu\nu} \partial_{\mu}\Phi\partial_{\nu}\Phi \right] \eqno(2)$$
where $H_{\mu\nu\lambda} = \partial_{\mu} B_{\nu\lambda} +$
cyclic perm.$R^{(D)}$ is the D-dimensional Ricci scalar and
$\Lambda$ is the cosmological constant equal to $\frac{(D -
26)}{3}$ for the bosonic string and $\frac{(D - 10)}{2}$ for the
fermionic string. It has been shown that [2] for (a) $X \equiv
\left(\hat{Y}^{m}, \tilde{Y}^{\alpha}\right), 1 \leq m \leq d, 1 \leq
\alpha \leq D-d$. $\hat{Y}^{m}$ having Euclidean signature, 
(b) background fields independent of $\hat{Y}^{m}$, and \\
(c) $G = \left(\matrix{
\hat{G}_{mn}	&0\cr
0		&\tilde{G}_{\alpha\beta}
\cr}\right) \enskip ; \enskip
B = \left(\matrix{
\hat{B}_{mn}	&0\cr
0		&\tilde{B}_{\alpha\beta}
\cr}\right)$ 
the action (1) can be recast into
$$S = - \int d^{D}\hat{Y}\int d^{D-d}\tilde{Y}\sqrt{det}\tilde{G}e^{-\chi}
\phantom{..........................}$$
$$[\Lambda -
\tilde{G}^{\alpha\beta}\tilde{\partial}_{\alpha}\chi
\tilde{\partial}_{\beta}\chi - \left({1 \over 8}\right)
\tilde{G}^{\alpha\beta} Tr \left(\tilde{\partial}_{\alpha} ML
\tilde{\partial}_{\beta} ML\right)
-\tilde{R}^{(D-d)}(G) + \left({1\over12}\right)\tilde{H}_{\alpha\beta\gamma}
\tilde{H}^{\alpha\beta\gamma}] \eqno(3)$$
where
$$L = \left(\matrix{
0	&1\cr
1	&0
\cr}\right) \eqno(4)$$
$$\chi = \Phi - ln \sqrt{det \hat{G}} \eqno(5)$$
$$M = \left(\matrix{
\hat{G}^{-1}		&- \hat{G}^{-1} \hat{B} \cr
\hat{B} \hat{G}^{-1}  	&\hat{G} - \hat{B} \hat{G}^{-1} \hat{B}
\cr}\right) \eqno(6)$$

If one of the coordinates $\hat{Y}^{1}$ is time-like, then the
action $(2)$ is invariant under an $O(d-1, 1) \otimes O(d-1, 1)$
transformation on $\hat{G}, \hat{B}$ and $\Phi$ given by
$$M \rightarrow \left({1 \over 4}\right) \left(\matrix{
\eta(S+R)\eta	&\eta(R-S)\cr
(R-S)\eta	&(S+R)
\cr}\right) 
M 
\left(\matrix{
\eta\left(S^{T}+R^{T}\right)\eta 	&\eta\left(R^{T}-S^{T}\right)\cr
\left(R^{T}-S^{T}\right)\eta		&\left(S^{T}+R^{T}\right)
\cr}\right) \eqno(7)$$
with $\eta$ = diag (-1, 1, ........1) ; S, R some O(d-1, 1)
matrices satisfying $S \eta S^{T} = \eta$,$ R \eta R^{T} = \eta$ ;
$S_{11} = cosh \theta = R_{11}$ ,$ S_{21} = -sinh \theta = -R_{21}$,
and $  S_{i1} = R_{i1} = 0 $,   for $ i \geq 3$.

In component form the transformed fields are given by [2] :
$$\left(\hat{G}^{'-1}\right)_{ij} = \left({1 \over
4}\right)[\eta(S+R)\eta \hat{G}^{-1}\eta\left(S^{T}+R^{T}\right)\eta$$
$$ + \eta(R-S) \left(\hat{G} -
\hat{B}\hat{G}^{-1}\hat{B}\right)\left(R^{T}-S^{T}\right)\eta$$
$$ -\eta(S+R)\eta \hat{G}^{-1}\hat{B} \left(R^{T}-S^{T}\right)\eta$$
$$ +\eta(R-S) \hat{B}\hat{G}^{-1}
\eta\left(S^{T}+R^{T}\right)\eta]_{ij} \eqno(8a)$$
$$\left(\hat{B'}\right)_{ij} = \left({1 \over 4}\right)[\{(R-S)\eta
\hat{G}^{-1}\eta \left(S^{T}+R^{T}\right)\eta$$
$$ + (S+R)\left(\hat{G} - \hat{B}\hat{G}^{-1}\hat{B}\right)
\left(R^{T}-S^{T}\right)\eta$$
$$ + (S+R) \hat{B} \hat{G}^{-1}\eta
\left(S^{T}+R^{T}\right)\eta$$
$$ - (R-S)\eta \hat{G}^{-1} \hat{B}
\left(R^{T}-S^{T}\right)\eta\}\hat{G'}]_{ij} \eqno(8b)$$
$$\Phi' = \Phi - \left({1\over 2}\right) ln\enskip\ det \hat{G} +
\left({1\over 2}\right) ln\enskip\ det \hat{G'} \eqno(8c)$$
The equations of motion obtained from (2) are
$$R_{\mu\nu} = D_{\mu}D_{\nu}\Phi + \left({1\over 4}\right)
H^{\lambda\rho}_{\mu} H_{\nu\lambda\rho} \eqno(9a)$$
$$D_{\mu}\Phi D^{\mu}\Phi - 2 D_{\mu}D^{\mu}\Phi + R -
\left({1\over 12}\right) H_{\mu\nu\rho} H^{\mu\nu\rho} = 0 \eqno(9b)$$
$$D_{\lambda} H^{\lambda}_{\mu\nu} - \left(D_{\lambda}
\Phi\right) H^{\lambda}_{\mu\nu} = 0 \eqno(9c)$$
Maximal symmetry implies
$$R_{\mu\nu} = K(1 - D) g_{\mu\nu} \enskip\ i.e.\enskip\ R = K(1 -D)D \eqno(10)$$

For $B_{\mu\nu} = 0$, equations $(9a)$  and $(10)$ give
$$D^{\mu}\phi D_{\mu}\phi = \partial^{\mu}\phi \partial_{\mu}\phi 
= R = K (1 - D) D \eqno(11a)$$
and $$ \partial^{\mu} \partial_{\nu}\phi = R_{\mu\nu} 
= K (1 - D) g_{\mu\nu} \eqno(11b)$$

Let $\phi$ be a function of $r$ only.This is a consistent assumption in the 
implementation of the O(d)$\otimes$O(d) symmetry (i.e. $r$ is the only $\tilde{Y}$
type coordinate while others are of $\hat{Y}$ type).Then $(11b)$ gives for 
$\mu = \nu = t,   {\partial}^2_{t} \phi = 0 $. Hence $\phi = \alpha r + \beta$,
where $\alpha, \beta $ are constants.This value of $ \phi $ when substituted 
in $(11a)$ gives $\alpha=0$.So the dilaton background $\phi$ is a constant
i.e. $\phi = \beta$.

For $B_{\mu\nu} \not= 0$, maximally symmetric solutions to (9)
are ( for $D = 3$)
$$H^{2}_{t1r} = \frac{K(1 - D)}{2} g_{tt}\enskip\ g_{11} \eqno(12a)$$
i.e.$$ B_{t1} = \left[\frac{K(1 - D)}{2}\right]^{1\over 2} \int
dr \left[g_{tt}\enskip\ g_{11}\right]^{1\over 2} \eqno(12b)$$
and$$ \partial^2_{r} \Phi = 0\enskip\ i.e.\enskip \Phi =
\alpha\enskip\ r + \beta \eqno(12c)$$

Now consider a class of solutions to $(9)$ of the form 
$$ds^{2} = - f(r) dt^{2} + \Sigma^{d-1}_{i=1} dx ^{1} dx^{1} + dr^{2}\eqno(13a)$$
$$\phi = \phi_{0} = \beta = constant, \enskip\ B_{\mu\nu} = 0  \eqno(13b)$$
Here $f(r)$ is some function of $(r)$.For simplicity, we take $D=3$ with 
coordinates $(t,x^1,r)$.The components of $R_{iklm}$ are 
$$R_{trtr}= (1/2)\partial^2_{r}f - (1/4) f^{-1} (\partial_{r}f)^2 \eqno(14a)$$
$$R_{t1t1}=R_{r1r1}= 0 \eqno(14b)$$
All other components are either vanishing or related to the above by symmetry 
(antisymmetry) properties  of  $R_{iklm}$. Maximal symmetry now means that 
$$R_{trtr} = (1/2)\partial^2_{r}f - (1/4) f^{-1} (\partial_{r}f)^2
                = - K f \eqno(15a)$$
$$R_{t1t1} = R_{r1r1} = 0 = - K 1 =  - K\eqno(15b)$$
Equations $(15)$ imply that the metric of the general form $(13a)$ is 
{\it maximally symmetric  only if  K = 0.} This agrees with  our  earlier 
conclusion that  $K=0$ for $B_{\mu\nu}=0$. In this case $f(r)=(r + const.)^2$.
Thus the metric  $(13a)$ and the Minkowski metric have the same  number  of
eigenvalues of each sign and same $K \enskip i.e.\enskip K=0$.Hence they are related by a 
coordinate diffeomorphism.

The twisted solutions obtained from $(13)$ , using $(8)$ are 
$$ \hat{G}' = \left(\matrix{-f/[1+(1-f)sinh^{2}\theta] & 0\cr
0 & 1/[1+(1-f)sinh^{2}\theta]
\cr}\right) = \left(\matrix{-E_{t} & 0\cr
0 & E_{1}
\cr}\right)\eqno(16)$$         

$$\hat{B}'=\left((1-f)cosh\theta sinh\theta/[1+(1-f)sinh^{2}\theta]\right)\enskip 
\left(\matrix{0 & 1\cr
-1 & 0
\cr}\right)\eqno(17)$$

$$\phi' = \beta - ln [ 1 + ( 1 - f ) sinh^{2}\theta ] \eqno(18)$$
The O(d)$\otimes$O(d) twisted metric is thus given by 
$$ds^{2} = - E_{t} dt^{2} + E_{1} dx^{1} dx^{1} + dr^{2} \eqno(19)$$
The components of the curvature tensor are now
$$R_{t1t1} = (1/4)\enskip \partial_{r} E_{t}\enskip \partial_{r}E_{1} \eqno(20a)$$
$$R_{trtr} = (1/2) \enskip \partial_{r}^{2}E_{t}\enskip  -\enskip  (1/4) \enskip (E_{t})^{-1}\enskip(\partial_{r}E_{t})^{2}\eqno(20b)$$
$$R_{1r1r} = (1/2) \enskip \partial_{r}^{2}E_{1}\enskip  +\enskip  (1/4) \enskip (E_{1})^{-1}\enskip(\partial_{r}E_{1})^{2}\eqno(20c)$$
All other components are either vanishing or related to the above by symmetry 
(antisymmetry) properties of $R_{iklm}$.The condition of maximal symmetry
now gives 
$$ y  z    =  -  K' \eqno(21a)$$
$$\partial_{r} y  +  y^{2}  =  - K'\eqno(21b)$$
$$\partial_{r} z  +  z^{2}  =  - K'\eqno(21c)$$
where 
$$y = \partial_{r}[(1/2) ln E_{t}] , \enskip  z = \partial_{r}[(1/2) ln E_{1}]\eqno(21d)$$   
and $K'$  is the new curvature constant.The twisted solution $(17)$ has 
$\hat{B}_{\mu\nu}' \ne 0 $.However, we have seen from the equations of motion
that maximal symmetry demands that (from eqation $(12b)$)
$$ B_{t1}' = [K' (1 - D)/2]^{1/2} \int dr\enskip [g_{tt}'\enskip g_{11}']^{1/2}$$
which when solved for $B_{t1}'$ results  in
$$B_{t1}' = - \left( (K')^{1/2}/2 sinh^{2}\theta \right) ln[1+(1-(r+const.)^{2}) sinh^{2}\theta]\eqno(22)$$
The solution $(17)$ (with $f(r)=(r+const.)^{2}$) and the solution $(22)$ can 
never be matched to be identical for any value of the parameter $\theta$.So
maximal symmetry is destroyed.

Keeping $B_{\mu\nu}$ still zero, assume (for generality) that the following
constitute a solution to the equations of motion 
$$ds^{2}  =  - f_{t}(r) dt^{2} + f_{1}(r) (dx^{1})^{2} + dr^{2}\eqno(23a)$$
$$\phi = \phi_{0}  = \beta , \enskip  \enskip  \enskip B_{\mu\nu}=0\eqno(23b)$$
Here, unlike $(19)$, $f_{t}$ and $f_{1}$ are two independent functions of $r$.
The metric $(23a)$ is maximally symmetric when 
$$p \enskip \enskip  q  = - K_{1} \eqno(24a)$$
$$\partial_{r} p + p^{2} = - K_{1} \eqno(24b)$$
$$\partial_{r} q + q^{2} = - K_{1} \eqno(24c)$$
with 
$$p = \partial_{r}[(1/2) ln f_{t}], \enskip \enskip \enskip q= \partial_{r}[(1/2) ln f_{1}]\eqno(24d)$$
and the solutions to $f_{t}(r)$ and $f_{1}(r)$ are found to be 
$$f_{t}(r)= cos^{2}(K_{1}^{1/2} r),\enskip \enskip \enskip f_{1}(r) = sin^{2}(K_{1}^{1/2} r),\enskip \enskip \enskip K_{1} > 0\eqno(25a)$$
$$f_{t}(r)= cosh^{2}(K_{1}^{1/2} r),\enskip \enskip \enskip f_{1}(r) = sinh^{2}(K_{1}^{1/2} r),\enskip \enskip \enskip K_{1} < 0\eqno(25b)$$
We shall now see that these solutions are inconsistent with the equations of
motion.

In the discussion immediately after $(11b)$ we saw that for $B_{\mu\nu}=0$ 
consistency of the the equations $(9)$ demands that the curvature constant vanishes.
Therefore $K_{1}=0$. This combined with $(24a)$ means that either $p=0$ or$q=0$,
i.e. either $f_{t}$ or $f_{1}$ must be a constant. Let $f_{1}=1$. Then $(23a)$
reduces to the form $(13a)$ which we have already discussed.We therefore conclude 
that for a string moving in a background of constant $\phi$ and vanishing
$B_{\mu\nu}$ , a maximally symmetric space-time solution is only possible with
a zero curvature metric. The O(d)$\otimes$O(d) twist , however, violates this 
maximal symmetry.
                    
Now consider the case of $B_{\mu\nu} \not= 0$. We take the
metric as in $(23a)$
$$ds^{2} = - f_{t}(r) dt^{2} + f_{1}(r) \left(dx_{1}\right)^{2} + dr^{2} \eqno(26a)$$
$$f_{t}(r) = cos^{2}\left(K_{1}^{1/2} r\right),\enskip  f_{1}(r) =
sin^{2}\left(K_{1}^{1/2} r\right),\enskip   K_{1} > 0 \eqno(26b)$$
$$f_{t}(r) = cosh^{2}\left(K_{1}^{1/2} r\right),\enskip  f_{1}(r) =
sinh^{2}\left(K_{1}^{1/2} r\right),\enskip   K_{1} < 0 \eqno(26c)$$
and assume that the curvature is small. Why we assume this will
be evident shortly. The equations (12) imply
$$H^{2}_{t1r} = \frac{K(D - 1)}{2} f_{t} f_{1} \eqno(27)$$
$$and \enskip\enskip\partial^{2}_{r} \Phi = 0\enskip i.e.\enskip \Phi = \alpha r + \beta \eqno(28)$$
(9b) then yields 
$$\alpha^{2} = K(1 - D)D + \left({1\over 4}\right) K(1 - D) \eqno(29)$$
This means that $\alpha^{2}$ is of the order of K. The equation
of motion for $H_{\mu\nu\lambda}$ (i.e.(9c)) is satisfied for
the derived solutions (14) and (15) except when $\mu = 1, \nu =
0$ and this non-vanishing part is 
$$\alpha \left[\frac{K(D - 1)}{2} f_{t} f_{1}\right]^{1\over 2} \eqno(30)$$

For $B_{\mu\nu} \not= 0$, these solutions are valid and can be
made compatible with the equations of motion as follows. In the
light of$(25)$,$(28)$ and the assumption of small $K$, $(30)$ is of
the order of $K^{5\over 2}$. Retaining upto terms linear in the
curvature, the contribution $(29)$ may be ignored and the equations of motion
satisfied. So in this approximation of small curvature, we can
have maximally symmetric solutions with $f_{t}$ and $f_{1}$, in
conjunction with $\Phi = \alpha r + \beta$ and a non-vanishing
$B_{\mu\nu}$. We confine ourselves to the case of positive
curvature. Identical conclusions also hold for negative
curvature.$(25a$) and $(12b)$  give the solution for $B_{\mu\nu}$ as
$$\hat{B}_{t1} = - \left({1\over 4}\right) cos \left[2
K_{1}^{1/2} r\right] \eqno(31)$$
so our starting solution with a maximally symmetric metric is
(26), (28) and (31). Using (8), the twisted solutions are
$$\hat{G'} = \left(\matrix{
-F_{t}	&0\cr
0	&F_{1}
\cr}\right) \eqno(32a)$$

$$F_{t} = \frac{f_{t}}{\left[1 + \left(1 - f_{t}f_{1}\right) sinh^{2}\theta +
\hat{B}_{t1} \left(\hat{B}_{t1} sinh^{2}\theta + sinh 2\theta\right)\right]}$$
$$F_{1} = \frac{f_{1}}{\left[1 + \left(1 - f_{t}f_{1}\right) sinh^{2}\theta +
\hat{B}_{t1} \left(\hat{B}_{t1} sinh^{2}\theta + sinh 2\theta\right)\right]}$$
$$\hat{B'} = \frac{\left({1\over 2}\right)\left(1 - f_{t}f_{1} +
\hat{B}^{2}_{t1}\right) sinh 2\theta + \hat{B}_{t1} cosh
2\theta}{1 + \left(1 - f_{t}f_{1}\right) sinh^{2}\theta +
\hat{B}_{t1} \left(\hat{B}_{t1} sinh^{2}\theta + sinh
2\theta\right)} \left(\matrix{0		&1\cr	-1	&0
\cr}\right) \eqno(32b)$$
$$\Phi' = \alpha r + \beta - ln \left[1 + \left(1 -
f_{t}f_{1}\right) sinh^{2}\theta +
\hat{B}_{t1}\left(\hat{B}_{t1} sinh^{2}\theta + sinh
2\theta\right)\right] \eqno(32c)$$
The analogues of (12a) is now
$$H'^{\enskip 2}_{t1r} = K_{1}' \frac{(D - 1)}{2} F_{t} F_{1} \eqno(33)$$
and this may be solved to get the antisymmetric tensor field as :
$$\hat{B'}_{t1} = -
\left[\frac{2\left[\frac{K'_{1}}{K_{1}}\right]^{1\over
2}}{\left[sinh\theta \left(12 + 11 sinh^{2}\theta\right)^{1\over
2}\right]}\right]$$
$$tan^{-1}\left[\frac{\left(5 sinh\theta
cos(2K_{1}^{1/2} r) - 4 cosh \theta\right)}{2\left(12 + 11
sinh^{2}\theta\right)^{1\over 2}}\right] \eqno(34)$$
$(32b)$ and $(34)$ can never be matched to be identical for any
value of $\theta$. So maximal symmetry is not preserved under
the O(d) $\otimes$ O(d) transformation.

The implication of this work is therefore that in the context of maximally
symmetric classical solutions to the string equations of motion, the
O(d) $\otimes$ O(d) transformation gives solutions that are {\it not} maximally
symmetric. However, as torsion is present , it is worthwhile to investigate
the meaning of maximal symmetry in the presence of torsion.

\end{document}